\documentclass[usenatbib,useAMS,usedcolumn]{mnras}

\usepackage{newtxtext,newtxmath}

\usepackage[T1]{fontenc}

\DeclareRobustCommand{\VAN}[3]{#2}
\let\VANthebibliography\thebibliography
\def\thebibliography{\DeclareRobustCommand{\VAN}[3]{##3}\VANthebibliography}

\usepackage{graphicx}	
\graphicspath{{fig/}}
\usepackage{amstext}
\usepackage{siunitx}
\usepackage{mathtools}
\usepackage{color}
\usepackage{twoopt}
\usepackage{hyperref} 
\usepackage{afterpage}
\usepackage{silence}
\usepackage{hyperref}

\let\sun\odot
\DeclareSIUnit\lightspeed{$c$}
\DeclareSIUnit\rydberg{Ry}
\DeclareSIUnit\erg{erg}
\DeclareSIUnit\magnitude{mag}
\DeclareSIUnit\jansky{Jy}
\DeclareSIUnit\gauss{G}
\DeclareSIUnit\h{$h$}
\DeclareSIUnit\hseven{$h$_7}
\DeclareSIUnit\parsec{pc}
\DeclareSIUnit\year{yr}
\DeclareSIUnit\solarluminosity{\ensuremath{L_\sun}}
\DeclareSIUnit\solarmass{\ensuremath{M_\sun}}
\DeclareSIUnit\solarmassinenergy{\ensuremath{M_\sun|c^2}}
\DeclareSIUnit\solarradius{\ensuremath{R_\sun}}
\DeclareSIUnit\arcsecond{as}
\DeclareSIUnit\astronomicalunit{au}
\DeclareSIUnit\clight{\ensuremath c}

\bibpunct{(}{)}{;}{a}{}{,}             
\makeatletter
  \newcommandtwoopt{\citeads}[3][][]{\href{http://adsabs.harvard.edu/abs/#3}%
    {\def\hyper@linkstart##1##2{}%
     \let\hyper@linkend\@empty\citealp[#1][#2]{#3}}}
  \newcommandtwoopt{\citepads}[3][][]{\href{http://adsabs.harvard.edu/abs/#3}%
    {\def\hyper@linkstart##1##2{}%
     \let\hyper@linkend\@empty\citep[#1][#2]{#3}}}
  \newcommandtwoopt{\citetads}[3][][]{\href{http://adsabs.harvard.edu/abs/#3}%
    {\def\hyper@linkstart##1##2{}%
     \let\hyper@linkend\@empty\citet[#1][#2]{#3}}}
  \newcommandtwoopt{\citeyearads}[3][][]%
    {\href{http://adsabs.harvard.edu/abs/#3}
    {\def\hyper@linkstart##1##2{}%
     \let\hyper@linkend\@empty\citeyear[#1][#2]{#3}}}
\makeatother
\hypersetup{pdfpagemode = {UseNone},
            pdftitle = {The dispersal of protoplanetary discs},
            pdfcreator = {\LaTeX},
            pdfproducer = {pdfeTeX-0.\the\pdftexversion\pdftexrevision},
            pdfauthor = {Giovanni Picogna, Barbara Ercolano, Catherine Espaillat},
            pdfsubject = {},
            pdfview = {FitH},
            pdfstartview = {FitH},
            colorlinks = {true},
            linkcolor = [rgb]{0,0.35,0.7},
            citecolor = [rgb]{0,0.35,0.7},
            filecolor = [rgb]{0.61,0,0},
            urlcolor = [rgb]{0,0.35,0.7},
           }

\defcitealias{2019MNRAS.487..691P}{Paper~I}
\defcitealias{2021MNRAS.tmp.2430E}{Paper~II}

\title[The dispersal of protoplanetary discs]{The dispersal of protoplanetary discs. -- III: Influence of stellar mass on disc photoevaporation}

\author[G. Picogna et al.]{
Giovanni Picogna,$^{1}$ \href{https://orcid.org/0000-0003-3754-1639}{\includegraphics[scale=0.04]{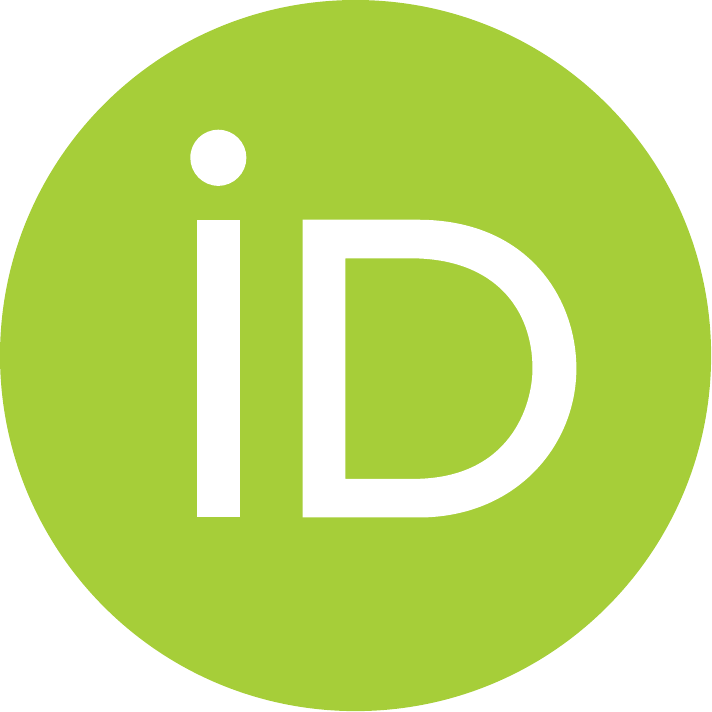}} \thanks{E-mail: picogna@usm.lmu.de}
Ercolano Barbara,$^{1,2}$ \href{https://orcid.org/0000-0001-7868-2740}{\includegraphics[scale=0.04]{orcid}}
Catherine~C. Espaillat$^{3}$ \href{https://orcid.org/0000-0001-9227-5949}{\includegraphics[scale=0.04]{orcid}}
\\
$^{1}$Universit\"{a}ts-Sternwarte, Ludwig-Maximilians-Universit\"{a}t M\"{u}nchen,
        Scheinerstr. 1, D-81679 M\"{u}nchen, Germany\\
$^{2}$Excellence Cluster Origins, Boltzmannstrasse 2, D-85748 Garching bei M\"{u}nchen, Germany\\
$^{3}$Department of Astronomy \& Institute for Astrophysical Research, Boston University, 725 Commonwealth Avenue, Boston, MA 02215, USA
}

\date{Accepted XXX. Received YYY; in original form ZZZ}

\pubyear{2021}

\begin{document}
\label{firstpage}
\pagerange{\pageref{firstpage}--\pageref{lastpage}}
\maketitle

\begin{abstract}
The strong X-ray irradiation from young solar-type stars may play a crucial role in the thermodynamics and chemistry of circumstellar discs, driving their evolution in the last stages of disc dispersal as well as shaping the atmospheres of newborn planets.
In this paper we study the influence of stellar mass on circumstellar disc mass-loss rates due to X-ray irradiation, extending our previous study of the mass-loss rate's dependence on the X-ray luminosity and spectrum hardness. We focus on stars with masses between $0.1$ and \SI{1}{\solarmass}, which are the main target of current and future missions to find potentially habitable planets.
We find a linear relationship between the mass-loss rates and the stellar masses when changing the X-ray luminosity accordingly with the stellar mass.
This linear increase is observed also when the X-ray luminosity is kept fixed because of the lower disc aspect ratio which allows the X-ray irradiation to reach larger radii.
We provide new analytical relations for the mass-loss rates and profiles of photoevaporative winds as a function of the stellar mass that can be used in disc and planet population synthesis models.
Our photoevaporative models correctly predict the observed trend of inner-disc lifetime as a function of stellar mass with an increased steepness for stars smaller than \SI{0.3}{\solarmass}, indicating that X-ray photoevaporation is a good candidate to explain the observed disc dispersal process.
\end{abstract}




\section{Introduction}

Our understanding of the physical processes driving the evolution of protoplanetary discs and ultimately the formation of planets is based on the observational evidence of disc lifetimes.
\citetads{1989AJ.....97.1451S} found that accretion discs are ubiquitous around young stellar objects (\SI{\sim 1}{Myr}). The initial disc fraction is almost independent of stellar mass (from $0.1$ up to \SI{10}{\solarmass}) and stellar environment (\citeads{2000AJ....120.3162L}, \citeads{2006A&A...451..177B}).
After \SI{10}{Myr} the picture changes completely since more than \SI{90}{\percent} of stars show no emission within \SI{1}{\astronomicalunit} \citepads{2004ApJ...612..496M}, and emission from small grains is found only in a few per cent of discs (\citeads{2004ApJ...608..526L}, \citeads{2005AJ....129.1049C}, \citeads{2006ApJ...639.1138S}).
This sets an upper limit to the disc dispersal process within $10$ to \SI{20}{Myr} (\citeads{2007ApJ...671.1784H}, \citeads{2010A&A...510A..72F}, \citeads{2014A&A...561A..54R}). Moreover, fitting the fraction of discs with dust emission at different wavelengths, \citetads{2014A&A...561A..54R} found an e-folding time of \SIrange{2}{3}{Myr} at \SIrange[]{3}{12}{\micro\meter} and \SIrange[]{4}{6}{\mega\year} at \SIrange[]{22}{24}{\micro\meter}, hinting at an inside-out dispersal of proto-planetary discs or to a second generation dust production \citepads[see also][]{2013MNRAS.428.3327K}.
The dust disc lifetime as a function of the stellar mass is less well characterised but the dispersal timescale appears to be up to two times faster for high mass stars ($\geq$ \SI{2}{\solarmass}, \citeads{2015A&A...576A..52R}).
The mass accretion onto the central star has also been studied extensively, though its evolution is less well constrained. Nevertheless, it has been shown that the characteristic timescale of disc accretion is shorter than that of dust disc dispersal (\citeads{2006ApJ...648.1206J}, \citeads{2010A&A...510A..72F}), and the accretion rate falls off as a function of time with a power law (e.g., \citeads{2012ApJ...755..154M}, \citeads{2014A&A...572A..62A}, \citeads{2016ARA&A..54..135H}).

The main physical process driving disc dispersal is still largely unconstrained, and it might change during the disc evolution and at different locations in the disc. A large body of evidence is pointing at magnetic disc winds as the main mechanism responsible for angular momentum transfer and mass-loss \citepads{2016ApJ...821...80B}, particularly at early times. Photoevaporative disc winds may also co-exist at early times and perhaps dominate at later stages (\citeads{2017RSOS....470114E}; \citeads{2020MNRAS.496..223W}). From a theoretical standpoint, there is a push to obtain better models that can be linked to observables to test their validity.

In the first paper of this series (\citeads{2019MNRAS.487..691P}, hereafter \citetalias{2019MNRAS.487..691P}) we showed the dependence between the stellar X-ray luminosity and the mass-loss rate due to thermal winds generated by the XEUV heating from the central star. Then we studied the influence of carbon depletion \citepads{2019MNRAS.490.5596W} and stellar spectra hardness on the mass-loss rates (\citeads{2021MNRAS.tmp.2430E}, hereafter \citetalias{2021MNRAS.tmp.2430E}). In this following work, we focus on stellar mass dependence and how it might influence disc evolution.

In Section~\ref{sec:methods} we briefly discuss the numerical set-up adopted, following our previous work. We then describe the main results in Section~\ref{sec:results} and discuss their theoretical and observational implications in Section~\ref{sec:discussion}. The main conclusions are then drawn in Section~\ref{sec:conclusions}.

\section{Methods}\label{sec:methods}
We run a series of hydrodynamical simulations following the approach outlined in \citetalias{2019MNRAS.487..691P} that we briefly describe here for completeness.

\begin{figure*}
    \centering
    \includegraphics[width=\textwidth]{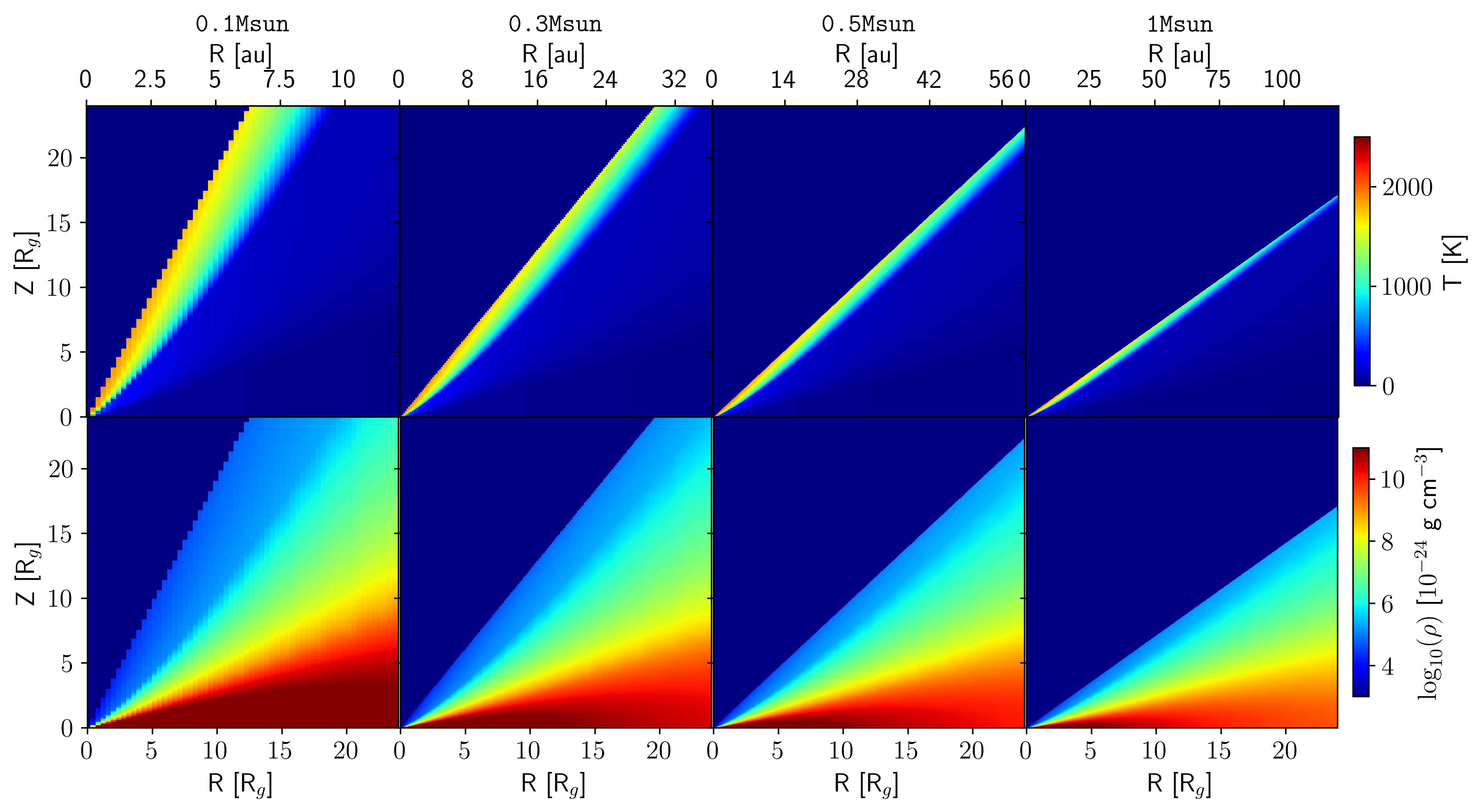}
    \caption{Initial dust temperature (upper panel) and density (lower panel) distribution in the inner $25$ gravitational radii for the the different runs from the \textsc{diad} models. \label{fig:initdiscs}}
\end{figure*}

The initial set-up is based on the gas densities and dust temperatures from the hydrostatic disc models from the D’Alessio Irradiated Accretion Disk (\textsc{diad}) radiative transfer models \citepads{1998ApJ...500..411D,1999ApJ...527..893D,2001ApJ...553..321D,2005ApJ...621..461D,2006ApJ...638..314D}, that best fit the median spectral energy distribution (SED) in Taurus. 
For this particular study a different set-up is adopted for each individual stellar mass, based on its mass and bolometric luminosity (see Figure~\ref{fig:initdiscs}).
The stellar parameters were obtained from \citetads{2000A&A...358..593S} for an age of \SI{1}{Myr}, and a metallicity $Z=0.02$ without overshooting.

We then ran the gas photoionization and dust radiative transfer code \textsc{mocassin} \citepads{2003MNRAS.340.1136E,2005MNRAS.362.1038E,2008ApJS..175..534E}, that solves the heating and cooling terms for various physical and irradiation properties at thermal equilibrium, to obtain a temperature prescription in the upper layers of the discs when irradiated by an X(EUV) stellar spectrum.
We adopted X-ray luminosities scaled as a function of stellar mass following \citetads{2007A&A...468..353G}
\begin{equation}\label{eq:Lx}
	\log_{10}{(\mathrm{L}_X)} = (1.54 \pm 0.12) \log_{10}{(\mathrm{M}_\star)} + (30.31 \mp 0.06)\,,
\end{equation}
although a recent analysis by \citetads{2021A&A...648A.121F} found a steeper dependence.

\begin{figure}
    \includegraphics[width=\columnwidth]{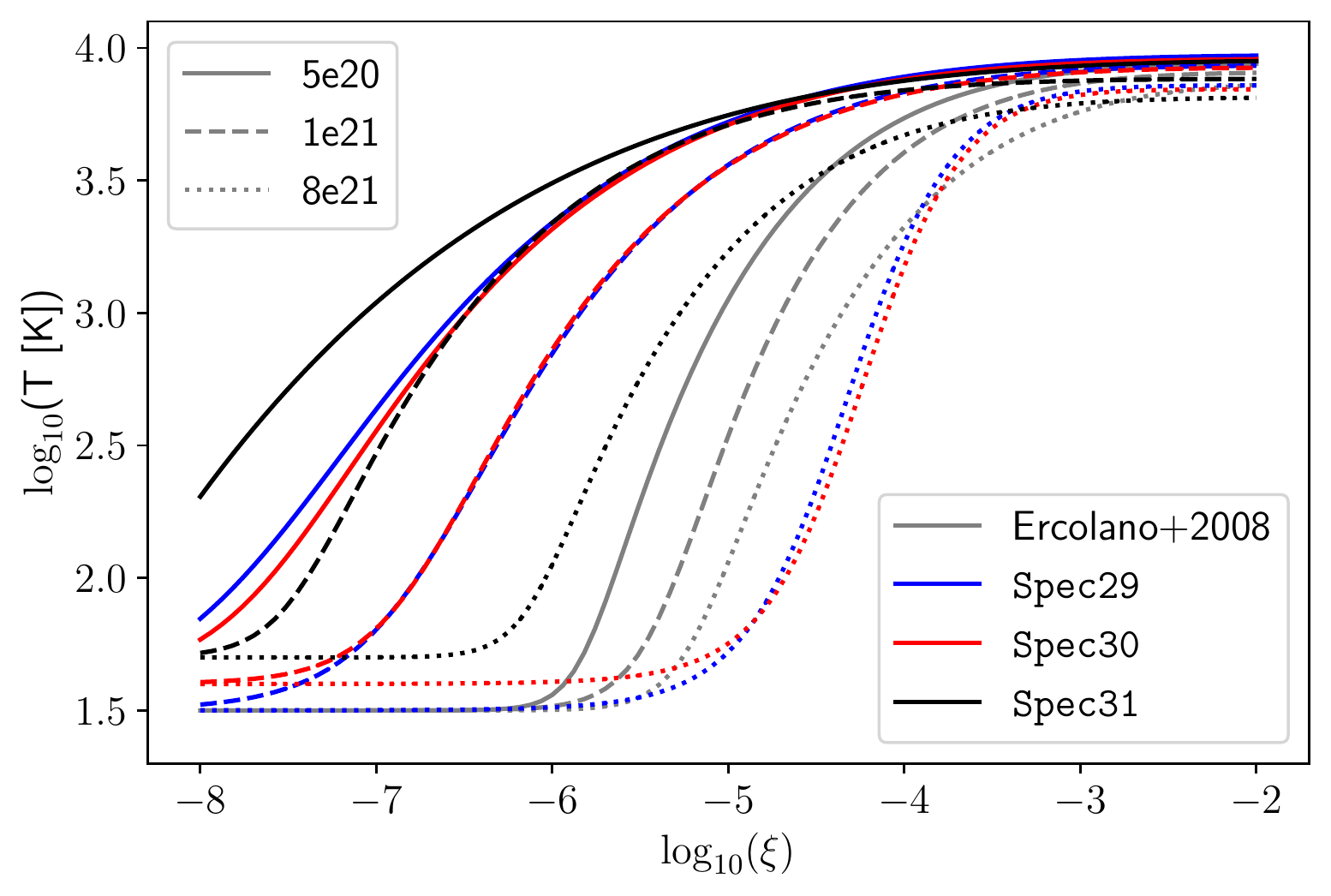}
    \caption{Temperature as a function of the ionization parameter, where $3$ selected column densities are highlighted \citepalias{2021MNRAS.tmp.2430E}. Three different spectral hardness are shown based on the stellar X-ray luminosity, from \texttt{Spec29} corresponding to \SI{e29}{erg.s^{-1}} to \texttt{Spec31} equal to \SI{e31}{erg.s^{-1}}.} \label{fig:tempxi}
\end{figure}

The temperature prescription is shown in Figure~\ref{fig:tempxi} for the different spectral hardness (from \texttt{Spec29} corresponding to \SI{e29}{erg.s^{-1}} to \texttt{Spec31} equal to \SI{e31}{erg.s^{-1}}) following \citetalias{2021MNRAS.tmp.2430E}.
It relates the local gas temperature to the column density to the central star (from \SI{5e20}{pp.cm^{-2}} to \SI{2e22}{pp.cm^{-2}}), and the local ionization parameter \citepads{1969ApJ...156..943T}
\begin{equation}
    \xi = \frac{L_X}{n r^2}\,,
\end{equation}
where $L_X$ is the X-ray luminosity, $n$ the number density, and $r$ the spherical radius from the central star. We compared it for completeness with the previously derived temperature prescription (used in \citetalias{2019MNRAS.487..691P}) obtained for a star with $L_X=\SI{2e30}{erg.s^{-1}}$ in \citetads{2008ApJS..175..534E}.

Finally we ran a set of hydrodynamical simulations with a modified version of the \textsc{pluto} code \citepads{2007ApJS..170..228M} presented in \citetalias{2019MNRAS.487..691P}, in order to use the temperature prescription from \textsc{mocassin} for column densities lower than the maximum penetration depth of X-rays (\SI{\sim 2e22}{pp.cm^{-2}}), and a perfect coupling between gas and dust temperatures using the \textsc{diad} models for larger column densities.
We evolve the models until a steady state is reached for the disc structure and the gas streamlines in the wind.

We modelled $4$ different stellar masses, ranging from \SI{0.1}{\solarmass} to \SI{1}{\solarmass} which, together with the \SI{0.7}{\solarmass} studied in \citetalias{2019MNRAS.487..691P}, allow us to study in detail the stellar mass dependence on the mass-loss rates due to photoevaporative winds in T Tauri stars. The initial parameters adopted in the different runs are summarised in Table~\ref{tab:stars}.

\begin{table*}
\caption{Star and disc properties}
\label{tab:stars}
\centering
\begin{tabular}{c c c c c c c c c c}
\hline
Name & $\mathrm{M}_\star$ [\si{\solarmass}] & $\mathrm{R}_\star$ [\si{\solarradius}] & ST & $\mathrm{L}_\star$ [\si{\solarluminosity}] & $\mathrm{L}_X$ [\SI{e29}{erg.s^{-1}}] & $\mathrm{T}_\star$ [\si{\kelvin}] & M$_\mathrm{d}$ [\si{\solarmass}] & R$_\mathrm{in}$ [\si{\astronomicalunit}] & Spectrum\\
\hline
\hline
   \texttt{1Msun} & $1.0$ & $2.615$ & K6 & $2.335$ & $20.4$ & $4278$ & $0.0444$ & $0.445$ & $\texttt{Spec30}$\\
   \texttt{0.5Msun} & $0.5$ & $2.125$ & M1 & $0.9288$ & $7.02$ & $3771$ & $0.0363$ & $0.2225$ & $\texttt{Spec30}$\\
   \texttt{0.3Msun} & $0.3$ & $2.310$ & M5 & $0.6887$ & $3.20$ & $3360$ & $0.0292$ & $0.1335$ & $\texttt{Spec29}$\\
   \texttt{0.1Msun} & $0.1$ & $1.055$ & M6 & $0.0856$ & $0.59$ & $2928$ & $0.0264$ & $0.0445$ & $\texttt{Spec29}$\\
\hline
\end{tabular}
\end{table*}

\subsection{Hydrodynamical model}\label{sec:hydro-model}

We adopted a spherical coordinate system centred on the star.
The grid is logarithmically spaced in the radial direction with \SI{500}{cells}, in order to have better resolution in the inner region of the disc (R$_\mathrm{in}=$ \SI{0.5}{R_g}), where photoevaporation is mostly effective.
At the same time, it allows us to model the disc out to large radii (R$_\mathrm{out}$ = \SI{600}{\astronomicalunit}) without strongly increasing our computational costs and preventing boundary effects that can affect the stability of the wind flow.
For a detailed discussion of the numerical artefacts induced by the grid outer boundary, the reader is referred to \citetalias{2019MNRAS.487..691P}.
The grid is spaced linearly in the polar direction, with a refinement at the wind launching region (\SI{100}{cells} from $0.01$ to \SI{0.5}{rad}, \SI{200}{cells} from $0.5$ to \SI{1}{rad}, and $50$ cells from $1$ to $\pi / 2$ rad).

\begin{figure}
    \centering
    \includegraphics[width=\columnwidth]{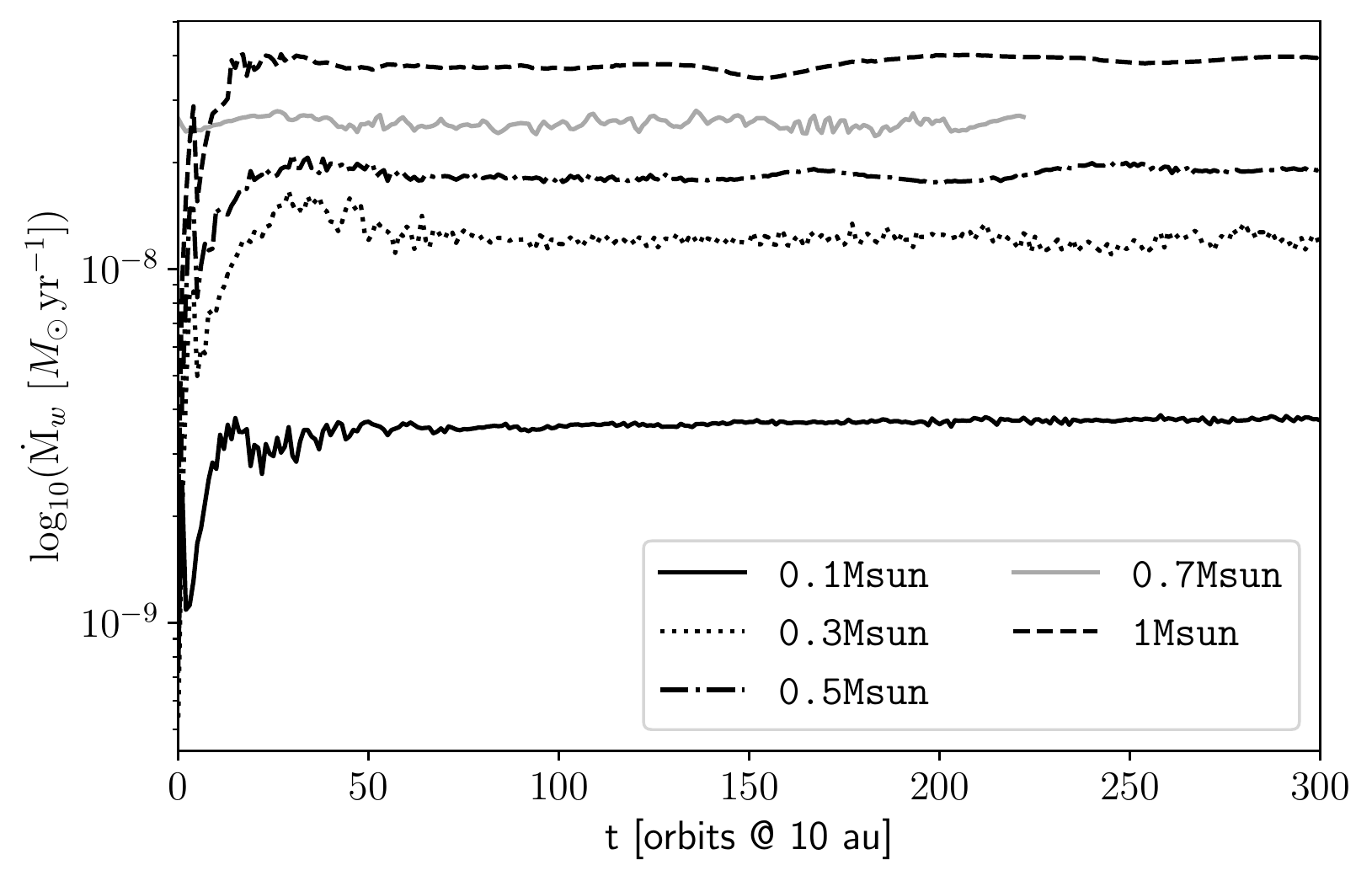}
    \caption{Cumulative mass-loss rate evolution as a function of time for the different stellar mass models. We included with a dark grey solid line the \SI{0.7}{\solarmass} run from \citetalias{2019MNRAS.487..691P}, rescaled to have the same X-ray luminosity from equation~
    \ref{eq:Lx}. \label{fig:mdotevol}}
\end{figure}

We initially evolve the system for few orbits at \SI{10}{au} without stellar irradiation in order for it to readjust to the hydrostatic equilibrium. Then, we switch on the stellar heating (applied through a parameterisation as explained above and in \citetalias{2019MNRAS.487..691P}) and continue to evolve the system for a few hundred orbits until the cumulative mass-loss rate and gas streamline in the wind have reached a steady-state after few hundred orbital periods (at \SI{10}{au}), as shown in Figure~\ref{fig:mdotevol}.

\section{Results}\label{sec:results}

There are two main locations that are of central importance in the study of photoevaporative winds.

(i) The gravitational radius $\mathrm{R}_g$, which defines the location where a gas parcel becomes unbound from the central star, equating the sound speed to the Keplerian speed (\citeads{1994ApJ...428..654H}, \citeads{2003PASA...20..337L}).
\begin{eqnarray}\label{eq:rg}
  \mathrm{R}_g &=& \frac{G \mathrm{M}_\star}{c_s^2} = \frac{\gamma-1}{2\gamma} \frac{G \mathrm{M}_\star \mu m_H}{k_B T_0} \nonumber \\ 
  &\simeq& 5.05 \left(\frac{T_0}{\SI{e4}{\kelvin}}\right)^{-1} \left(\frac{\mathrm{M}_\star}{\SI{1}{\solarmass}}\right) \ [\si{\astronomicalunit}] \,,
\end{eqnarray}
where $\gamma=5/3$ is the adiabatic index, $G$ the gravitational constant, $\mathrm{M}_\star$ the stellar mass, $\mu = 2.35$ the mean molecular weight, $m_H$ the proton mass, $k_B$ the Boltzmann constant, and $T_0$ is the temperature at the base of the flow, which is equal to \SI{e4}{\kelvin} for a pure EUV wind. 
\citetads{2012MNRAS.422.1880O} found analytically that the temperature in the flow should be fixed to first order by the stellar mass and thus be scale free when scaled by the gravitational radius.
Thus we will scale in our analysis always (when not differently stated) the winds by this value, in order to study differences in the expected behaviour. We have to keep in mind that, since we are not treating an isothermal wind, the gravitational radius will not be a fixed cylindrical radius but an area where the temperature will increase due to X-ray irradiation winning over the gravitational pull from the central star. Nevertheless for simplicity in the rest of the paper we will refer to $\mathrm{R}_g$ assuming a fixed temperature of \SI{e4}{K}, typical of a isothermal EUV wind.

(ii) The critical radius $\mathrm{R}_c$, which defines the location where the gas becomes supersonic, and for a Parker wind is 
\begin{equation}\label{eq:rcrit}
    \mathrm{R}_c = \frac{G M_\star}{2 c_s^2}\,,
\end{equation}
where $c_s$ is the gas sound speed. The main assumptions of a Parker wind are that the outflow is steady, spherically symmetric and isothermal. Each streamline of the thermal wind will pass through its critical radius, forming a sonic surface which is fundamental in order to understand the properties of the photoevaporative wind. When we have the density and sound speed at the sonic surface we can determine the mass-flux in the entire flow as mass-flux is conserved (\citeads{1983ApJ...271...70B}, \citeads{2012MNRAS.422.1880O}).

In this section we first discuss the effects of the irradiation from different stellar masses (and relative X-ray luminosities) on the global mass loss rates and the radial mass loss profiles obtained at steady state from our 2D hydrodynamic calculations. In the second part we isolate the effect of changing only the stellar mass (keeping fixed the X-ray luminosity) and compare it with previous studies.

\subsection{Disc profiles}
\begin{figure*}
    \centering
    \includegraphics[width=\textwidth]{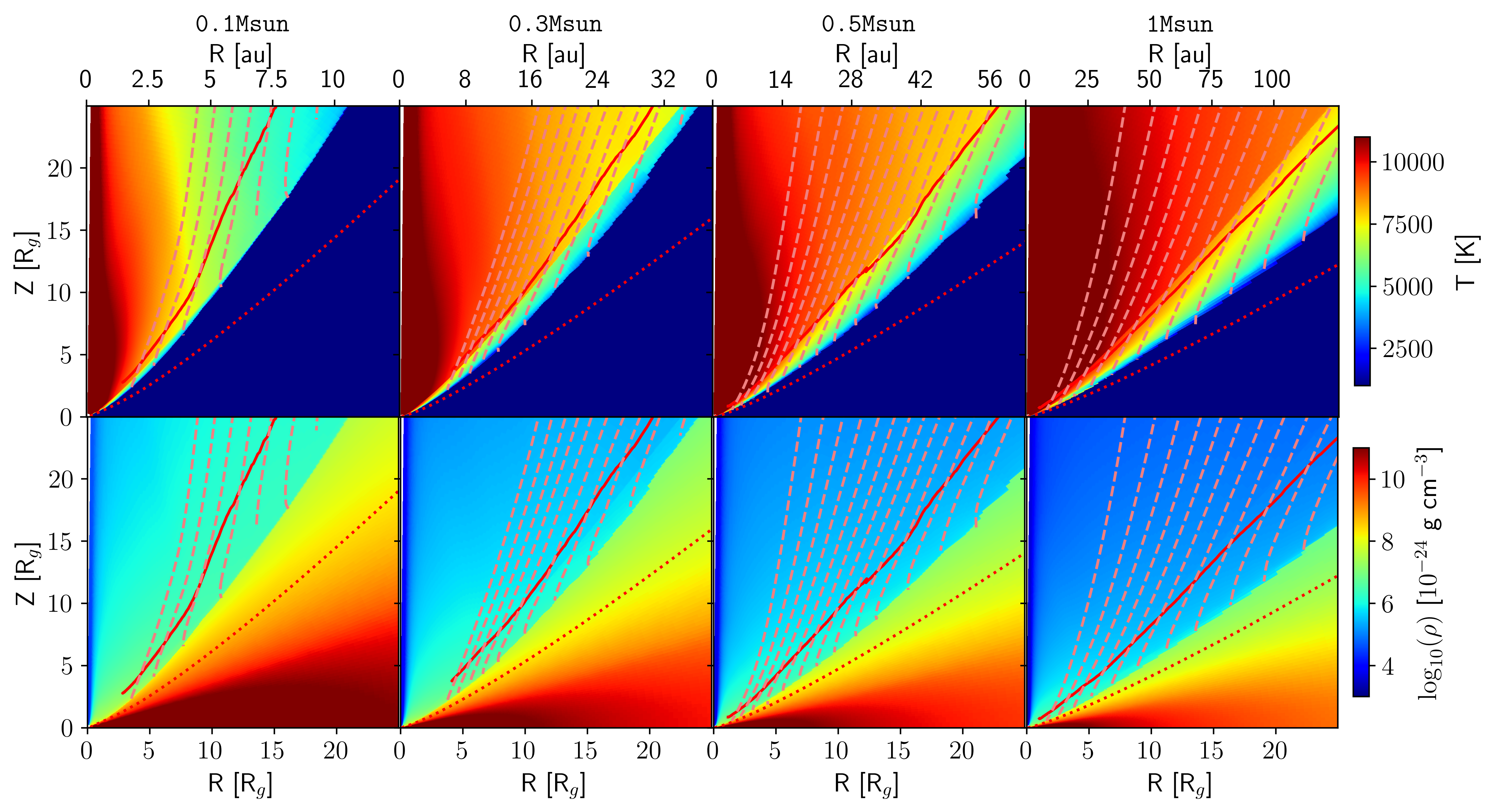}
    \caption{Temperature (upper panel) and density (lower panel) distribution in the inner $25$ gravitational radii for the the different runs. The sonic surface is overlayed with a solid red line and the gas streamlines every $5\%$ of the cumulative mass-loss rate with orange dashed lines. The maximum penetration depth of the X-rays is shown as well with a dashed red line. \label{fig:discs}}
\end{figure*}

We show in Figure~\ref{fig:discs} the gas temperature (top panel) and density (bottom panel) distribution for the different stellar masses at equilibrium.
\begin{figure}
    \centering
    \includegraphics[width=\columnwidth]{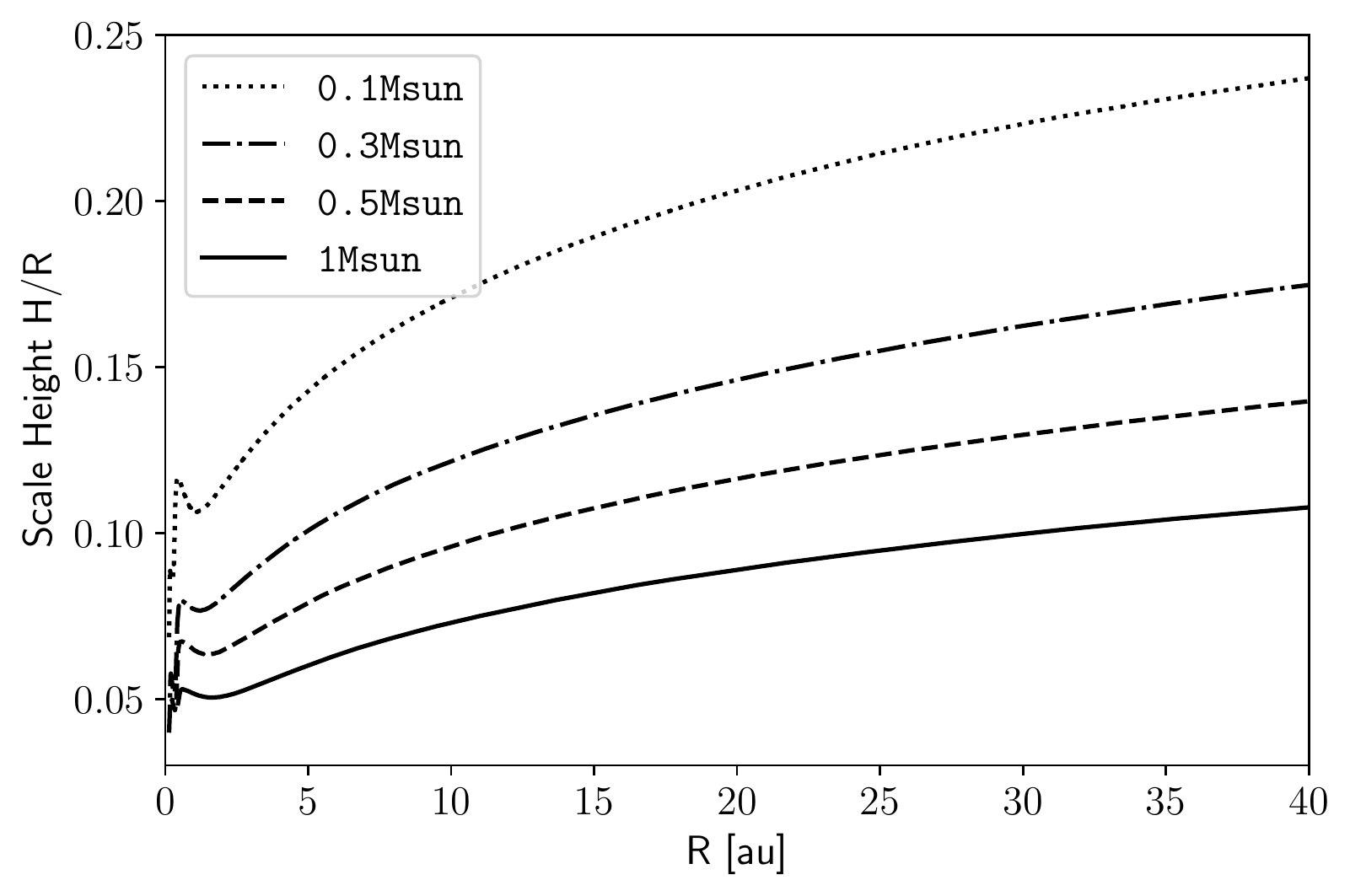}
    \caption{Initial disc aspect ratio for the runs reported in Tab.~\ref{tab:stars}. \label{fig:aspectratio}}
\end{figure}
Two main aspects can be seen in the observed profiles. 
(i) The disc aspect ratio is strongly dependant on the stellar mass, and it is not affected by X-ray photoevaporation (comparing the bottom panels of Figures \ref{fig:discs} and \ref{fig:initdiscs}), as the X-rays cannot reach the deeper dense regions of the disc until the mass accretion rate becomes comparable to the photoevaporative mass-loss rate. We show this in a more quantitative way in Figure~\ref{fig:aspectratio}, where we plot the disc aspect ratio as a function of radius for the different stellar masses. Within \SI{40}{R_g} the disc flaring changes from \SI{0.1}{\solarmass} with a disc aspect ratio of $h/r = 0.096 \cdot r^{0.25}$ to \SI{1}{\solarmass} with $h/r = 0.036 \cdot r^{0.3}$. The larger disc flaring for low stellar masses affects the vertical location of the maximum penetration depth of X-rays (shown in Figure~\ref{fig:discs} as a red dashed line) which shifts further away from the disc mid-plane.
(ii) The temperature structure in the wind region, even when rescaled by the gravitational radius, varies considerably for the different stellar masses, in contrast to what is predicted analytically in \citetads{2012MNRAS.422.1880O}, showing warmer winds for larger stellar masses and X-ray luminosities.

\subsection{Mass-loss rates}\label{sec:mdot}
\begin{figure}
  \includegraphics[width=\columnwidth]{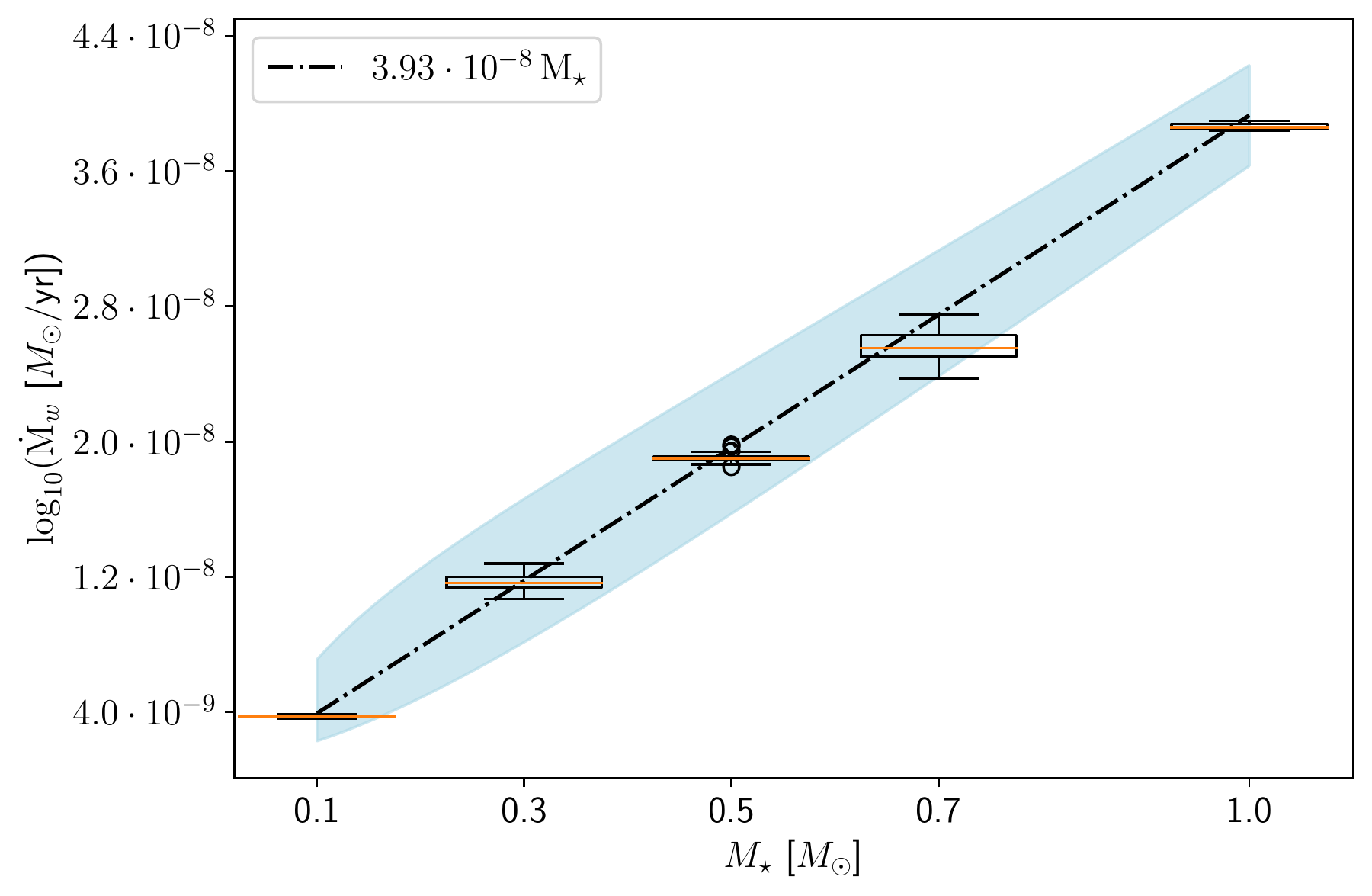}
  \caption{Box plot of the cumulative mass-loss rate as a function of stellar mass over the last $50$ orbital timescales at \SI{10}{\astronomicalunit}. The derived linear fit is overplotted with a black dashed line, and a blue shaded region shows the variation in the mass-loss rate due to the uncertainty in the X-ray luminosity as a function of stellar mass in equation~\ref{eq:Lx}. \label{fig:Mdot}}
\end{figure}

We integrated the gas flow along the streamlines in the wind during the last \SI{50}{orbits} of each simulation and plotted the resulting cumulative mass-loss rates as a function of stellar masses in a square-box plot in Figure~\ref{fig:Mdot}, where we included also the run with a \SI{0.7}{\solarmass} star from \citetalias{2019MNRAS.487..691P}, which has been rescaled in order to have the same stellar mass--X-ray luminosity adopted in these runs following equation~\ref{eq:Lx}.
A fitting function has been overplotted with a dashed line, which shows the linear dependence of the mass-loss rate as a function of stellar mass. This linear trend is in contrast with the flat distribution as a function of stellar mass ($\propto \mathrm{M}_\star^{-0.068}$) obtained by \citetads{2012MNRAS.422.1880O}, but there the X-ray luminosity was kept constant while changing the stellar mass (see Section~\ref{sec:mass-dependance} for a detailed comparison). Furthermore, the linear trend does not show the flattening of the mass-loss rates observed for large X-ray luminosity in \citetalias{2019MNRAS.487..691P} because even the \SI{1}{\solarmass} star has an X-ray luminosity of $\log(L_X) = 30.3$, which is smaller than the saturation value observed in the previous study.
The best fitting function obtained from the current study is
\begin{equation}\label{eq:MdotLxMass}
  \dot{\mathrm{M}}_w = 3.93\times10^{-8} \left(\frac{\mathrm{M}_\star}{\si{\solarmass}}\right) [\si{\solarmass.yr^{-1}}]\,.
\end{equation}

\subsection{Wind profiles}\label{sec:wind-prof}
\begin{figure}
  \centering
  \includegraphics[width=\columnwidth]{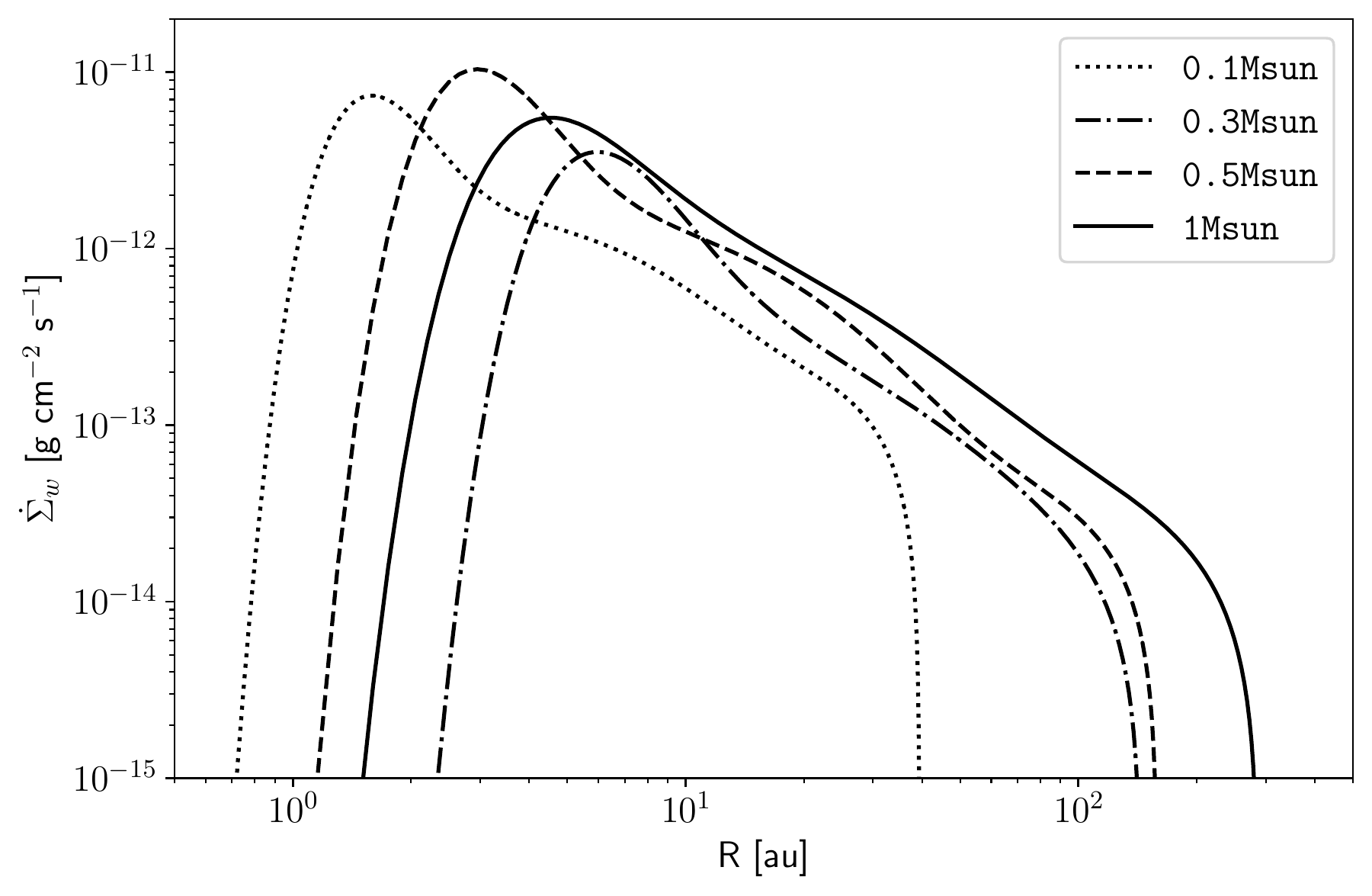}
  \includegraphics[width=\columnwidth]{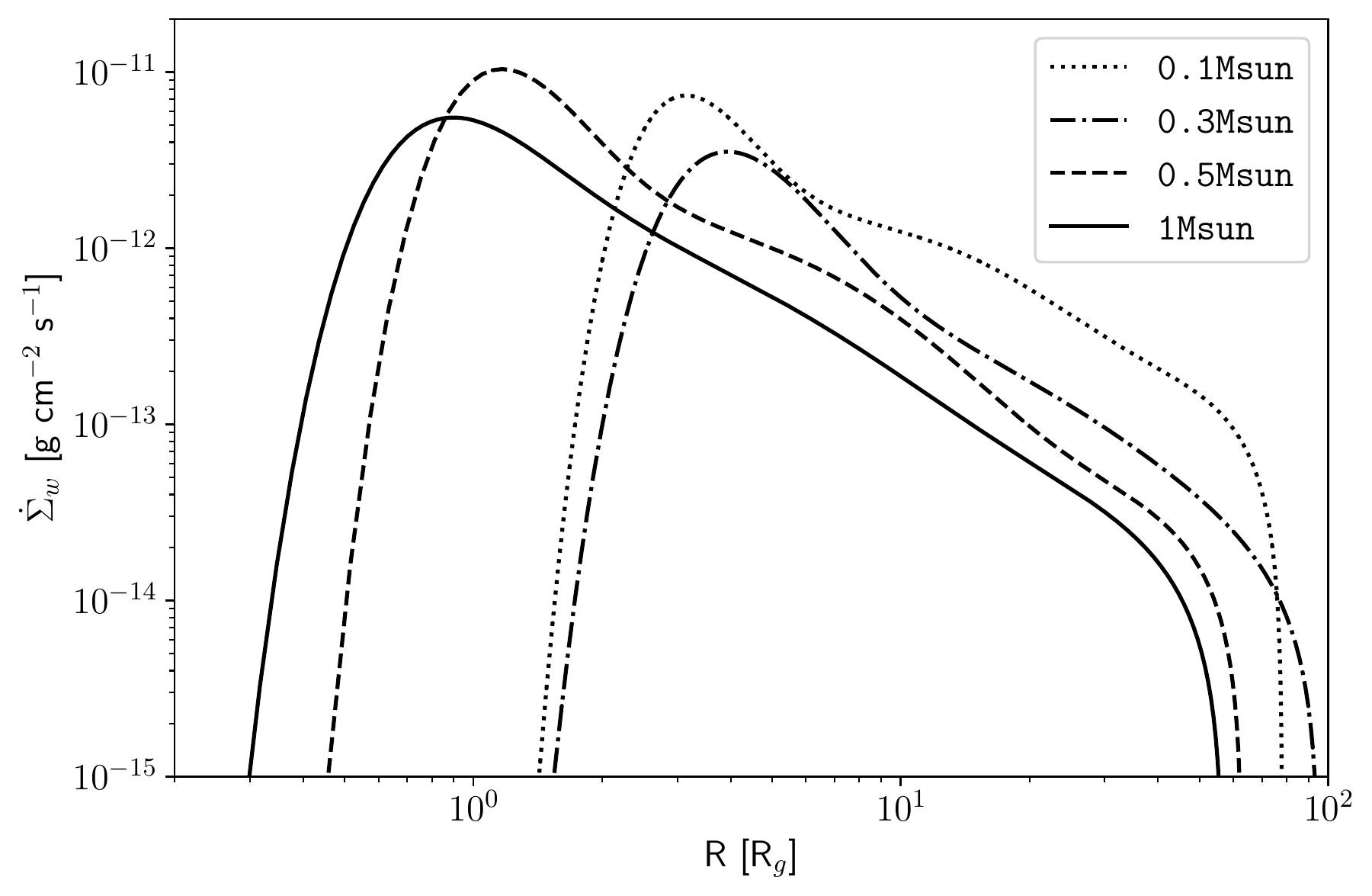}
  \caption{Surface density mass-loss rate profile for the different simulations. In the top panel the mass-loss rate is plotted as a function of the cylindrical radius in au, while in the bottom panel the dependence is a function of the gravitational radius (see section~\ref{sec:wind-prof}). \label{fig:Sigmadot}}
\end{figure}

From the derivative of the integrated mass-loss profile along the cylindrical radius we obtain the radial distribution of the mass-loss rate shown in Figure~\ref{fig:Sigmadot}. Looking at the top panel, two trends are visible. There is an increase of the peak location as a function of stellar mass, with the only exception for the \SI{0.3}{\solarmass} star. A similar increase is seen in the maximum reach of the wind in the outer regions of the protoplanetary disc, which is able to remove material up to \SI{\sim 35}{\astronomicalunit} for a \SI{0.1}{\solarmass} star, while it reaches \SI{\sim 270}{\astronomicalunit} for a \SI{1}{\solarmass} star. This is a direct result of the lower flaring for disc orbiting larger mass stars, which allows the wind to have a further reach in the outer disc regions and generate a more massive wind.

To understand the anomaly observed for the \SI{0.3}{\solarmass} star we rescaled the cylindrical radius by the gravitational radius in the bottom panel of Figure~\ref{fig:Sigmadot}.
We can see that, when rescaled, we obtain 2 similar profiles for the low X-ray luminosities (and spectra hardness) and two for the high luminosities. This can be explained with the harder spectra in the 'soft' X-ray band (0.1–1 keV) for lower X-ray luminosities which drives the peak at larger radii, as studied in \citetalias{2021MNRAS.tmp.2430E}. The relative difference amongst the profiles, once rescaled, is given by a combined effect of the change in the stellar mass and absolute value of the X-ray luminosity.

The best fit for the surface mass-loss rate is given by the following function, as in \citetalias{2019MNRAS.487..691P}
\begin{eqnarray}
  \label{eq:surf1}
  \dot{\Sigma}_w(R) &= \ln{(10)} \bigg(\frac{6\, a\, \ln{(R)}^5}{R\, \ln{(10)}^6} +
  \frac{5\, b\, \ln{(R)}^4}{R\, \ln{(10)}^5} +
  \frac{4\, c\, \ln{(R)}^3}{R\, \ln{(10)}^4} + \\ \nonumber
  &\frac{3\, d\, \ln{(R)}^2}{R\, \ln{(10)}^3} +
  \frac{2\, e\, \ln{(R)}}{R\, \ln{(10)}^2} + \\ \nonumber
  &\frac{f}{R\, \ln{(10)}}\bigg)
  \frac{\dot{\mathrm{M}}_w(R)}{2\pi\, R} \ [\SI{}{\solarmass.\astronomicalunit^{-2}.yr^{-1}}]\,
\end{eqnarray}
where
\begin{equation}
\label{eq:surf2}
  \frac{\dot{\mathrm{M}}_w(R)}{\dot{\mathrm{M}}_w(\mathrm{L}_X)} = 10^{a\log{R}^6 + b\log{R}^5 + c\log{R}^4 + d\log{R}^3 + e\log{R}^2 + f\log{R} + g}
\end{equation}
where the parameters for the different stellar masses are given in Table~\ref{tab:fit}.
\begin{table*}
\caption{Parameters for the Surface density profiles in equations~\ref{eq:surf1},\ref{eq:surf2}}
\label{tab:fit}
\centering
\begin{tabular}{c c c c c c c c c}
\hline
$\mathrm{M}_\star$ [\si{\solarmass}] & a & b & c & d & e & f & g & $\dot{\mathrm{M}}_w$ [\SI{e-8}{\solarmass.\year^{-1}}]\\
\hline
\hline
   $1.0$ & $-0.6344$ & $6.3587$ & $-26.1445$ & $56.4477$ & $-67.7403$ & $43.9212$ & $-13.2316$ & $3.86446$\\
   $0.5$ & $-1.2320$ & $10.8505$ & $-38.6939$ & $71.2489$ & $-71.4279$ & $37.8707$ & $-9.3508$ & $1.9046$\\
   $0.3$ & $-1.3206$ & $13.0475$ & $-53.6990$ & $117.6027$ & $-144.3769$ & $94.7854$ & $-26.7363$ & $1.17156$\\
   $0.1$ & $-3.8337$ & $22.9100$ & $-55.1282$ & $67.8919$ & $-45.0138$ & $16.2977$ & $-3.5426$ & $0.37588$\\
\hline
\end{tabular}
\end{table*}

\subsection{Stellar mass dependence}\label{sec:mass-dependance}
\begin{figure}
  \centering
  \includegraphics[width=\columnwidth]{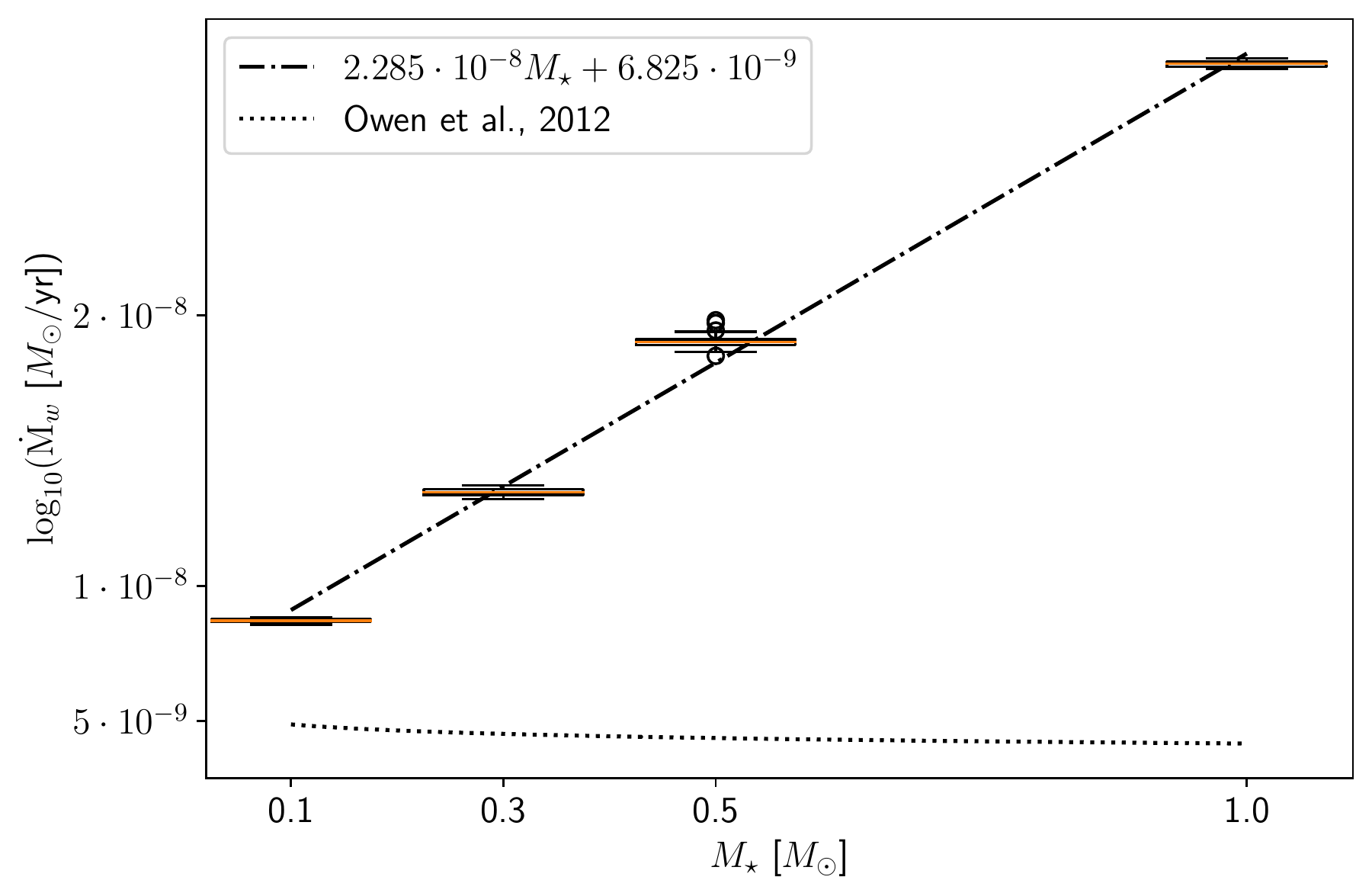}
  \caption{Box plot of the cumulative mass-loss rate as a function of stellar mass over the last 50 orbital timescales at \SI{10}{\astronomicalunit}, keeping the X-ray luminosity fixed. A linear fit has been overplotted with a dot-dashed line, as well as the prediction from \citepads{2012MNRAS.422.1880O} in a dotted line.\label{fig:cumdotMtest}}
\end{figure}

\begin{figure}
  \centering
  \includegraphics[width=\columnwidth]{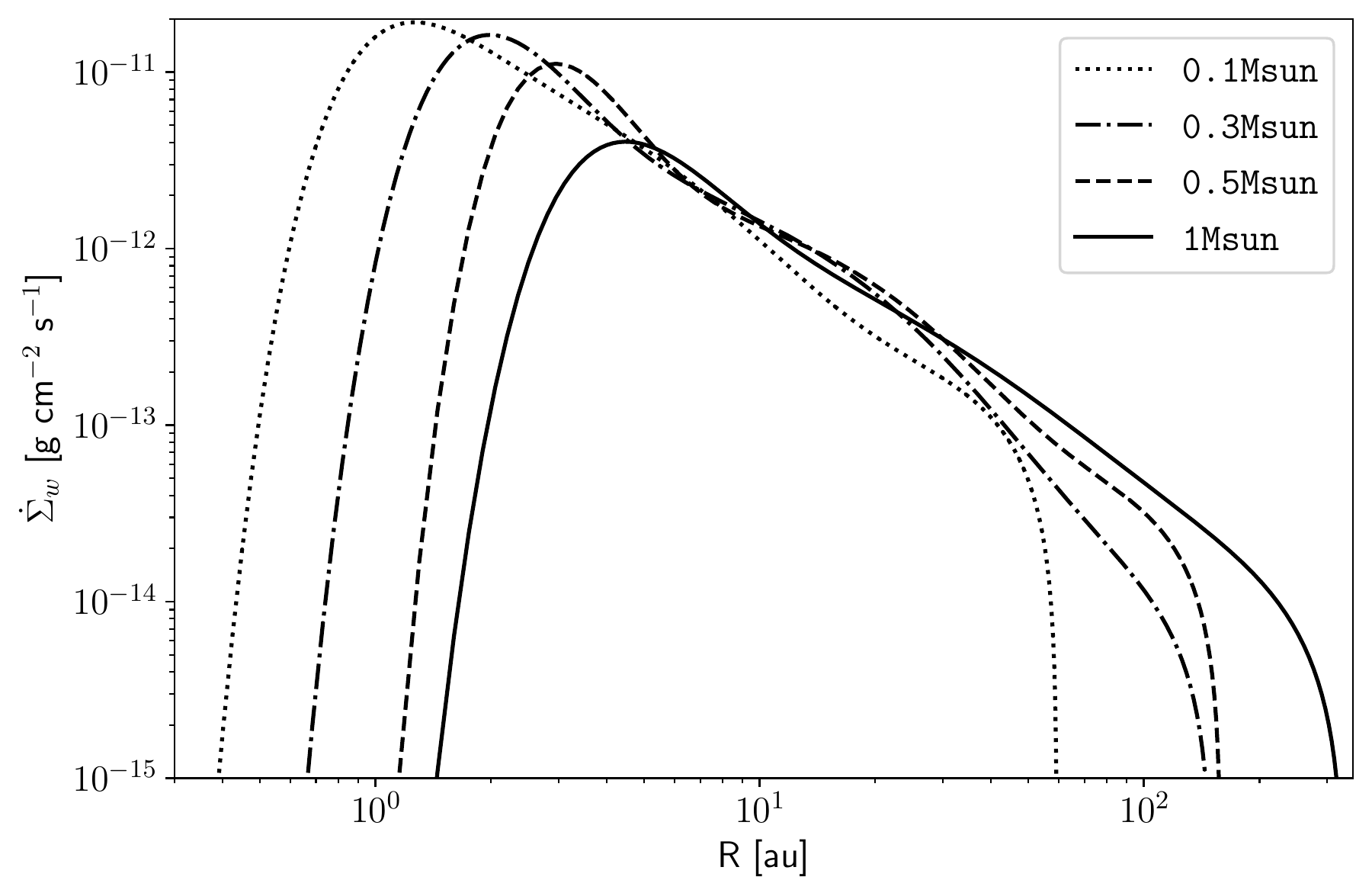}
  \caption{Surface density mass-loss rate profile as a function of cylindrical radius for the different simulations at fixed X-ray luminosity. \label{fig:SigmadotLx}}
\end{figure}

In order to isolate the effect of stellar mass on the photoevaporative mass-loss rates we performed a new set of simulations where we fixed the X-ray luminosity to the value of \SI{7e29}{erg.s^{-1}} (i.e. that of the \SI{0.5}{\solarmass} star).
We show in Figure~\ref{fig:cumdotMtest} the cumulative mass-loss rate for the different stellar masses and we compare it with the prescription from \citetads{2012MNRAS.422.1880O}.
We see that the mass-loss rate is not independent of the stellar mass, in contrast to what observed in \citetads{2012MNRAS.422.1880O}. However, the dependence on the stellar mass becomes much more shallow, increasing only by a factor $3$ from \SI{0.1}{\solarmass} to \SI{1}{\solarmass} instead of an order of magnitude as shown in Figure~\ref{fig:Mdot}. The linear fit that matches the data shown in Figure~\ref{fig:cumdotMtest} is
\begin{equation}
    \dot{\mathrm{M}}_w = 2.34 \times 10^{-8} \left(\frac{\mathrm{M}_\star}{\si{\solarmass}}\right) + 6.23 \times 10^{-9} [\si{\solarmass.yr^{-1}}]\,.
\end{equation}
To understand the origin of the stellar mass dependence we plotted in Figure~\ref{fig:SigmadotLx} the Surface mass-loss rate as a function of cylindrical radius for the different runs, which shows two main trends. An increase in the peak location and in the maximum reach as a function of stellar mass. 
The increased peak location is a direct consequence of the larger gravitational pull for more massive stars, which affects the location where eventually a gap will open in the disc. Although it scales linearly with stellar mass, it has little influence on the cumulative mass-loss rate since the disc mass scales as the square of the distance from the central star and the contribution from within the gravitational radius is negligible. Moreover \citetalias{2019MNRAS.487..691P} have shown that the cumulative mass-loss rate is insensitive to the inner hole radius for holes up to \SI{30}{\astronomicalunit} for a \SI{0.7}{\solarmass} star.
The increased reach of the stellar irradiation is the main effect driving up the mass-loss rates, since for intermediate cylindrical radii the profiles are lying one on top of each other. The reason why the runs with larger mass stars have a further reach is purely geometrical, since flatter discs allow the radiation to reach larger cylindrical radii before being absorbed. The change in the disc flaring as a function of stellar mass comes from the increase in the stellar bolometric luminosity (affecting the dust temperature) and in the disc mass from the DIAD models (see Table~\ref{tab:stars}).

Besides the change in the disc flaring, another difference between the runs (see Table~\ref{tab:stars}) is the disc mass. The bulk of the disc mass resides in the disc mid-plane which is not reachable by the stellar irradiation. However changing the disc mass affects the disc structure and thus the height at which the irradiation can penetrate. \citetads{2012MNRAS.422.1880O} found out that a variation in the disc mass was not changing the cumulative mass-loss rate, and more recently we confirmed this result in \citetads[][see their Figure~7]{2019MNRAS.490.5596W} where we saw a wind mass-loss rate independent of the disc mass for a large range of disc-to-star mass ratios and carbon abundances. We are thus confident that, at this stage of disc evolution, the disc mass is not affecting the cumulative wind mass-loss rate.

\section{Discussion}\label{sec:discussion}

In this section we discuss the differences between the properties at the sonic surface in our model in comparison with the theoretical prediction for a Parker wind and previous numerical results. We then compare the inner-disc lifetime obtained from our model with observationally derived one.

\subsection{Sonic surface}
\begin{figure}
    \centering
    \includegraphics[width=\columnwidth]{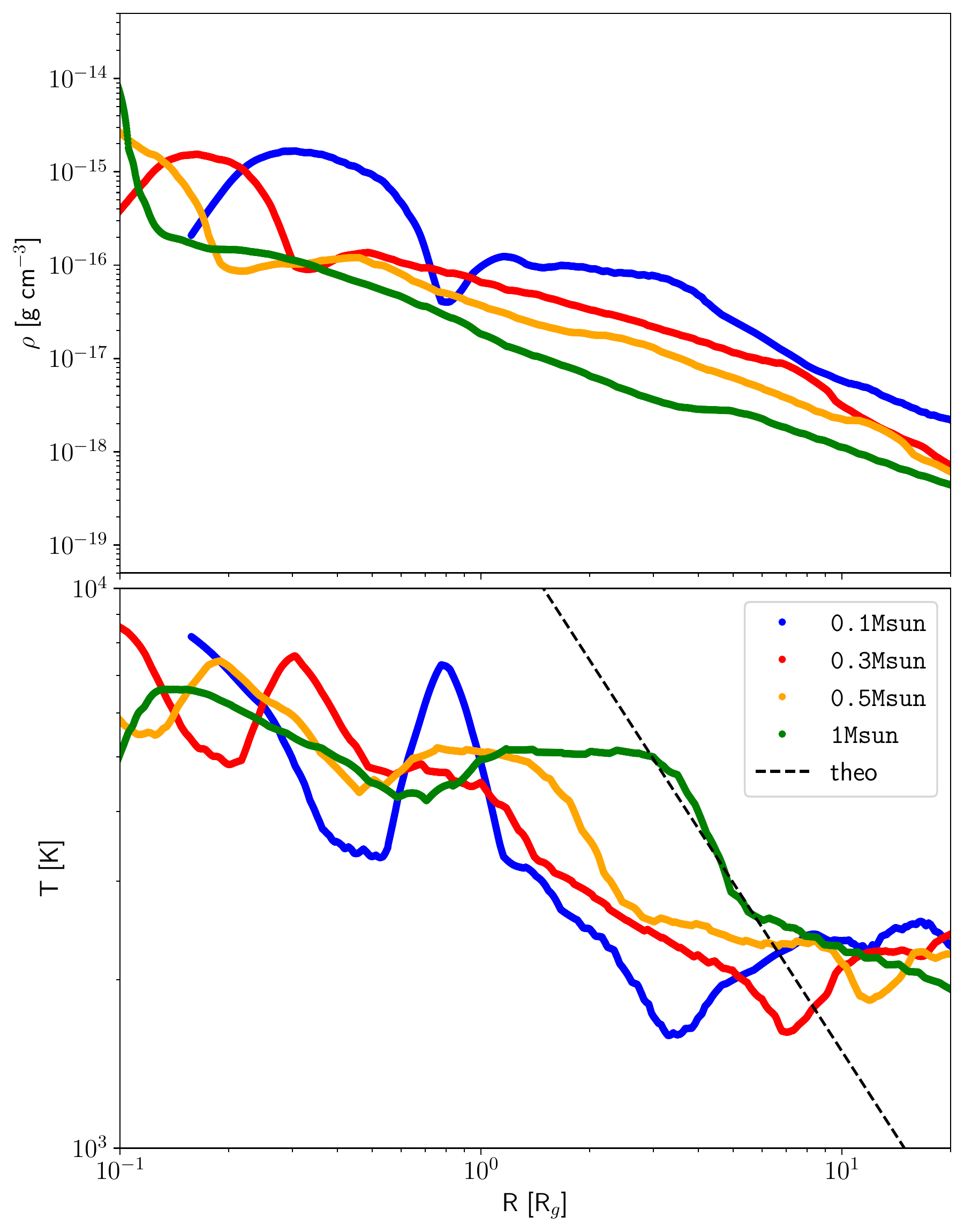}
    \caption{Temperature at the sonic surface as a function of the gravitational radius for the different stellar masses modelled. \label{fig:sonicsurf}}
\end{figure}

\begin{figure}
    \centering
    \includegraphics[width=\columnwidth]{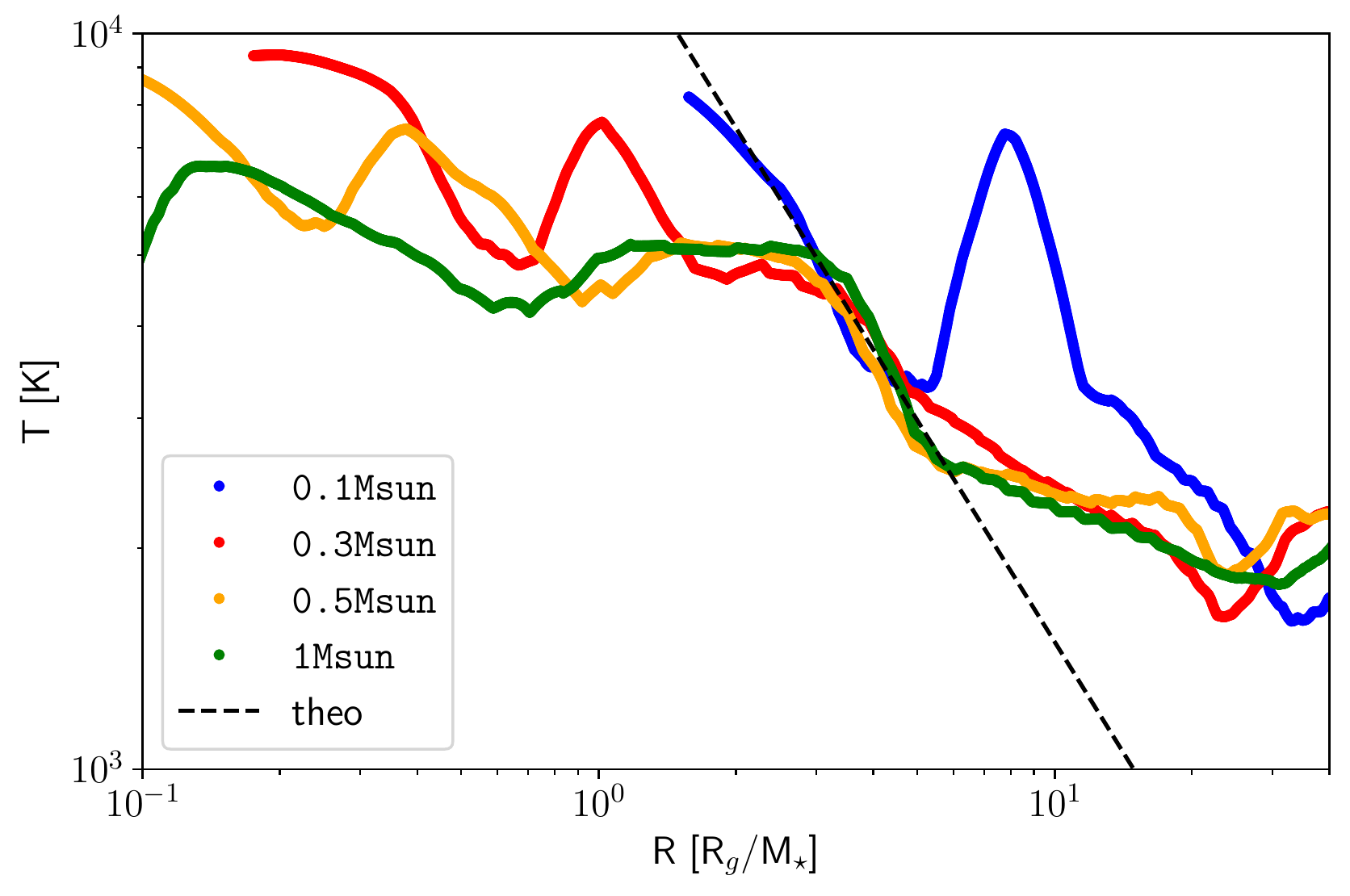}
    \caption{Temperature at sonic surface as a function of the gravitational radius divided by the stellar mass for the different models. \label{fig:sonicsurfrescaled}}
\end{figure}

The local gas properties at the sonic surface are fundamental to understand the overall wind properties.
For a disc wind, the sound speed at the sonic surface should be on the same order of magnitude of the Parker value which, from equation~\ref{eq:rcrit}, is
\begin{equation}
    {c_s}^2 \approx \frac{G \mathrm{M}_\star}{2R}\,.
\end{equation}
In Figure~\ref{fig:sonicsurf} we plot the density (upper plot) and temperature (bottom plot) at the sonic surface for the different runs, and we compare the temperature with the theoretical prediction for a Parker wind.
We see that in our case the steepness of the temperature profile is much shallow, and only the \SI{1}{\solarmass} star (green line) matches the expected profile between $3$ and $6$ gravitational radii before flattening again.
This behaviour has been observed also by \citeads{2012MNRAS.422.1880O} (see their Figure 6), where the Parker wind value was recovered only around the gravitational radius, flattening close to the star and far away from it, following the sigmoidal temperature prescription coming from Figure~\ref{fig:tempxi}.
Two trends can be identified in Figure~\ref{fig:sonicsurf}.
(i) The density at the sonic surface is almost linearly increasing with stellar mass, being \SI{9.2e-17}{g.cm^{-3}} for the \SI{0.1}{\solarmass} star and \SI{6.5e-18}{g.cm^{-3}} for \SI{1}{\solarmass}. The bump observed inside \SI{1}{R_g} is caused by the bump in the disc scale height from Figure~\ref{fig:aspectratio}, which comes directly from the \textsc{diad} models and is caused by an increase of the dust temperatures close to the star.
(ii) The temperature passes through a transition regime with a similar slope which is shifted towards the central star going from the higher to the lower mass star. This shift is also linearly increasing with the stellar mass, and to show it, we replot the bottom panel of Figure~\ref{fig:sonicsurf} dividing the gravitational radius again by the stellar mass in Figure~\ref{fig:sonicsurfrescaled}, where the theoretical temperature for the \SI{1}{\solarmass} star is reported to guide the eye.
The linear radial shift in the temperature at the sonic surface may be due to the difference in the disc scale height, which affects the density at the sonic surface, and thus the ionization parameter which is responsible for the change in the temperature. 

\subsection{Observational constraints}
The observed lifetimes of discs around stars of different masses presents a fundamental constraint for disc dispersal models. Our new models which span the observational range of observed stellar masses in disc surveys allow us to make the first such comparison for a disc-dispersal model.

We can define the inner-disc lifetime $t_\mathrm{life}$ as the time at which accretion rates $\dot{\mathrm{M}}_\mathrm{acc}$ decrease to photoevaporative rates $\dot{\mathrm{M}}_w$.
This happens because, at this time, a gap opens in the inner disc (around the gravitational radius) disconnecting the inner and outer disc evolution. 
The inner disc disperses on a short timescale (viscous timescale at few \si{\astronomicalunit}), not being replenished anymore by the outer disc.
The lifetime of the inner disc $t_\mathrm{life}$ can be approximated then by
\begin{equation}\label{eq:tlifedef}
    \dot{\mathrm{M}}_w = \dot{\mathrm{M}}_\mathrm{acc,0}\left(\frac{t_\mathrm{life}}{t_\nu}\right)^{-3/2}\, ,
\end{equation}
where $t_\nu$ is the viscous timescale
\begin{equation}
    t_\nu = \frac{M_\mathrm{disc,0}}{2 \dot{\mathrm{M}}_\mathrm{acc,0}}\, .
\end{equation}
Combining the previous equations we obtain
\begin{eqnarray}\label{eq:tlife}
    t_\mathrm{life} &=& t_\nu \left(\frac{\dot{\mathrm{M}}_\mathrm{acc,0}}{\dot{\mathrm{M}}_w}\right)^{2/3} = \frac{M_\mathrm{disc,0}}{2 \dot{\mathrm{M}}_\mathrm{acc}}\left(\frac{\dot{\mathrm{M}}_\mathrm{acc}}{\dot{\mathrm{M}}_w}\right)^{2/3} \nonumber \\
    &=& \frac{1}{2}\left(\frac{M_\mathrm{disc,0}}{\dot{\mathrm{M}}_\mathrm{acc,0}^{1/3}\,\dot{\mathrm{M}}_w^{2/3}}\right)\, ,
\end{eqnarray}
where the free parameters are given by the initial disc mass and accretion rates.

For several star forming regions an observational $\dot{\mathrm{M}}_\mathrm{acc}-\mathrm{M}_\star$ relation has been determined (\citeads{2017A&A...600A..20A}, \citeads{2017A&A...604A.127M}, \citeads{2021A&A...648A.121F}, \citeads{2021A&A...652A..72A}), where, like in the $1$ to \SI{3}{\mega\year} old Lupus, a sharp increase of the accretion rate is observed for low-mass stars ($<$\SI{0.2}{\solarmass}, \citeads{2017A&A...600A..20A})
\begin{equation}\label{eq:mdotAlcala}
    \log(\dot{\mathrm{M}}_\mathrm{acc}) = 
    \begin{cases}
      4.58 (\pm 0.68) \, \log(\mathrm{M}_\star) - 6.11(\mp 0.61), & \leq \SI{0.2}{\solarmass} \\
      1.37 (\pm 0.24) \, \log(\mathrm{M}_\star) - 8.46 (\mp 0.11), & \text{otherwise.}
    \end{cases}
\end{equation}
We can then use the current accretion rates to determine the initial accretion rate, assuming a constant disc viscous evolution
\begin{equation}\label{eq:mdot0}
    \dot{\mathrm{M}}_\mathrm{acc} = \dot{\mathrm{M}}_\mathrm{acc,0} \left(\frac{t}{t_\nu}\right)^{-3/2}\,,
\end{equation}
which leaves us only with the free parameter of the initial disc mass $M_\mathrm{disc,0}$.

An interesting star forming region to test the validity of our observational predictions is $\lambda$ Ori (also known as Collinder~$69$), since it has been extensively studied in X-ray emission \citepads{2011A&A...526A..21B}, and it seems to be at a critical age for disc dissipation (\SI{\sim 5}{\mega\year}).
\citeads{2012A&A...547A..80B} obtained the cumulative disc fraction as a function of stellar mass from infrared excess observations of this open cluster, and we can convert this into an approximate inner-disc lifetime, following the approach of \citeads{2021ApJ...910...51K} as
\begin{equation}
    t_\mathrm{life} \simeq -\frac{t_\mathrm{age}}{\ln{f_\mathrm{disc}}}\, ,
\end{equation}
where $t_\mathrm{age} = 5\, \mathrm{Myr}$ \citepads{2011A&A...536A..63B}.

\begin{figure}
    \centering
    \includegraphics[width=\columnwidth]{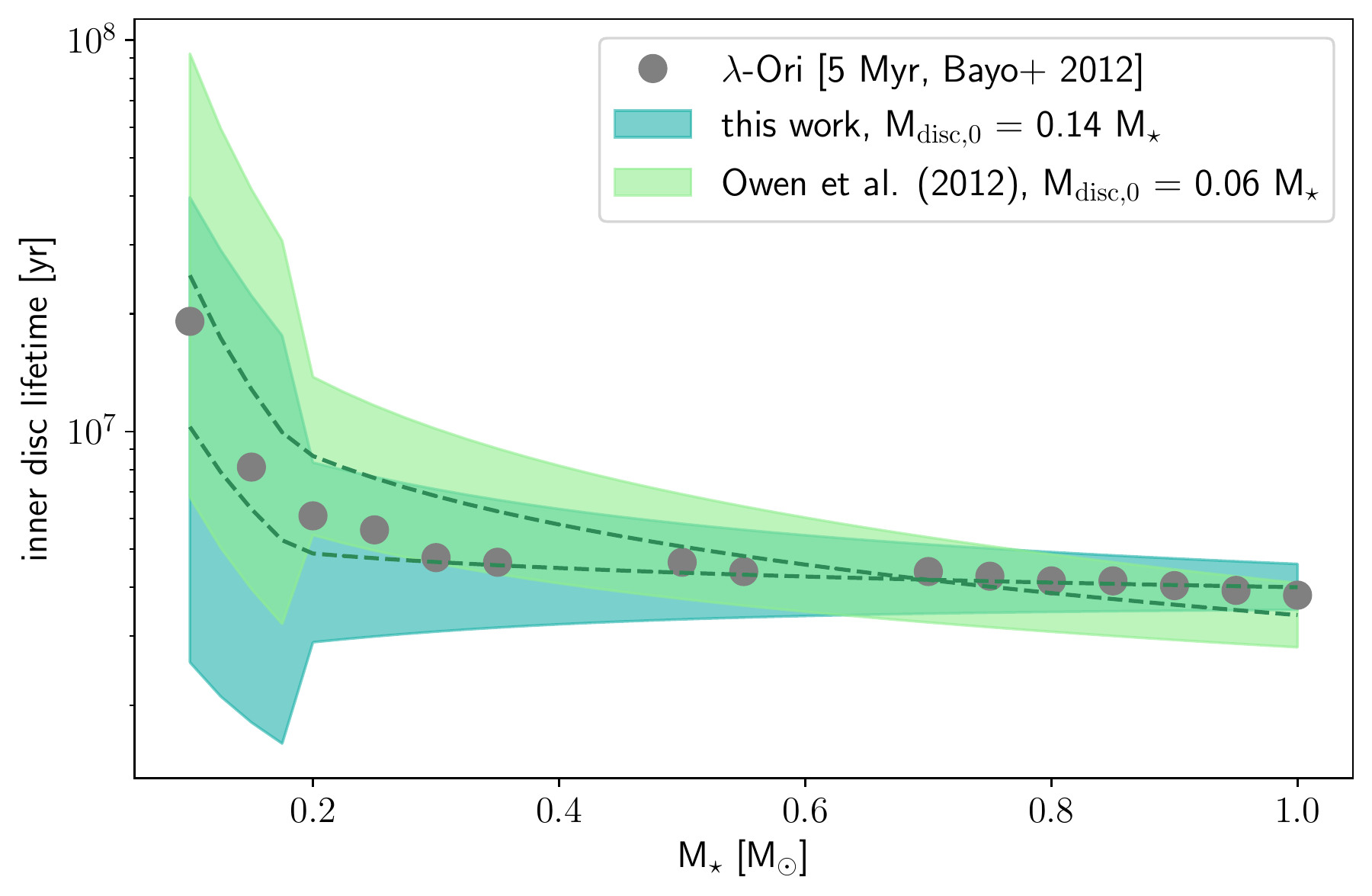}
    \caption{Inner-disc lifetime calculated from our models using equation~\ref{eq:tlife} compared with the disc fraction as a function of the mass for the members of $\lambda$ Ori. The shaded regions show the uncertainty in equations~\ref{eq:mdotAlcala},\ref{eq:Lx}. \label{fig:obs}}
\end{figure}

In Figure~\ref{fig:obs} we directly compared the observational data points derived from \citetads{2012A&A...547A..80B} in grey, with the inner-disc lifetime obtained from eqs.~\ref{eq:MdotLxMass}, \ref{eq:tlife} in sea-green and, for comparison, in light-green the result using the photoeavaporative mass-loss rate from equation~B1 of \citetads{2012MNRAS.422.1880O}. The shaded region represents the uncertainty in the observational relation between the accretion rates and stellar masses (equation~
\ref{eq:mdotAlcala}) and that in the X-ray luminosity as a function of stellar mass (equation~\ref{eq:Lx}). The sharp increase of the shaded regions for stars $<$ \SI{0.2}{\solarmass} is mainly caused by the large uncertainty of equation~\ref{eq:mdotAlcala} for small mass stars.

In order to get the best fit to the observation we used a linear relation between the initial disc mass and stellar mass and a initial value of \SI{0.14}{\mathrm{M}_\star} for the current models and \SI{0.06}{\mathrm{M}_\star} for \citetads{2012MNRAS.422.1880O}.
\citetads{2013ApJ...771..129A} found a roughly linear relationship between the disc mass and stellar mass, studying a large sample of Class~II discs. More recently \citetads{2016ApJ...831..125P}, studying a few young star forming regions, found a slightly steeper than linear dust disc mass - stellar mass relationship, though including in the sample also the brown dwarfs linearizes it again \citepads{2021arXiv210605247R}.
The results match very well the observations for the whole range of explored masses, and in particular the change in the slope for masses below \SI{0.3}{\solarmass}, which gives us confidence on the ability of the photoevaporative models to predict the observed inner-disc lifetimes. Furthermore, our new prescription, with a linear scaling of disc masses as a function of stellar mass, captures better the observed distribution with respect to the previous state of the art model.

The gap opening condition is local and it is usually satisfied in late stages of disc evolution inside \SI{10}{\astronomicalunit} (see e.g. \citetalias{2019MNRAS.487..691P}). Adopting the cumulative wind mass-loss rate in equation~\ref{eq:tlife} is a simplification, although the mass removed in the outer disc affects the mass flux reaching the inner disc. We are currently working to merge all our results in comprehensive disc and planet population synthesis models (Emsenhuber et al. in prep., K{\"u}ster et al. in prep.), but this simplified approach allows us nevertheless to test whether our models correctly predicts the observed behaviour.

\section{Conclusions}\label{sec:conclusions}

  We have computed for the first time comprehensive photoevaporation models spanning the whole observed range of low mass stars ($\leq$ \SI{1}{\solarmass}) using observationally derived X-ray stellar spectra. 
  We found that

   \begin{enumerate}
      \item stellar mass and bolometric luminosity have a strong impact on the underlying disc structure, changing its aspect ratio (see Figures~\ref{fig:initdiscs},\ref{fig:aspectratio}). 
      As a result, on the one hand the X-ray irradiation can penetrate deeper into the disc for smaller mass stars, generating denser and colder winds (see Figures~\ref{fig:discs},\ref{fig:sonicsurf}). On the other hand, the flatter discs around high mass stars allow the stellar irradiation to reach larger radii and generate more massive winds (see Figure~\ref{fig:mdotevol});
      \item we provide fitting functions in Table~\ref{tab:fit} for the surface mass-loss rate profiles for different stellar masses whose luminosity is scaled based on equation~\ref{eq:Lx} that can be readily applied to disc and planet population synthesis codes;
      \item the resulting cumulative mass loss rate scales linearly with the stellar mass (see equation~\ref{eq:MdotLxMass});
      \item the temperature as a function of the gravitational radius at the sonic surface scales linearly with the stellar mass (see Figure~\ref{fig:sonicsurfrescaled});
      \item the observed inner-disc lifetimes as a function of stellar mass can be very well reproduced by our new models, indicating that X-ray photoevaporation dominates the final phases of protoplanetary disc evolution (see Figure~\ref{fig:obs}).
   \end{enumerate}

\section{Data availability}
   The data underlying this article and the scripts used to create the Figures are available at \href{https://cutt.ly/lElY9JI}{https://cutt.ly/lElY9JI}.

\section*{Acknowledgements}

    We thank the anonymous referee for a thorough report and insightful suggestions which improved the paper.
    GP acknowledges support from the DFG Research Unit ‘Transition Disks’ (FOR 2634/2).
    This work was performed on the computing facilities of the Computational Center for Particle and Astrophysics (C2PAP).
    GP would like to thank Kristina Monsch and Tommaso Grassi for their insightful comments on the manuscript and during the development of the project.
    This research was supported by the Excellence Cluster ORIGINS which is funded by the Deutsche Forschungsgemeinschaft (DFG, German Research Foundation) under Germany's Excellence Strategy - EXC-2094 - 390783311.
    This paper utilizes the D’Alessio Irradiated Accretion Disk (DIAD) code. We wish to recognize the work of Paola D’Alessio, who passed away in 2013. Her legacy and pioneering work live on through her substantial contributions to the field.

\bibliographystyle{mnras}
\bibliography{biblio}

\bsp	
\label{lastpage}
\end{document}